\documentclass{article}

\usepackage{arxiv}
\usepackage[utf8]{inputenc}

\usepackage{longtable}
\usepackage{authblk}
\usepackage[caption = false]{subfig}
\usepackage{graphicx}
\usepackage{amssymb}
\usepackage{amsmath}
\usepackage{amsthm}
\usepackage[noend]{algpseudocode}
\usepackage{algorithm}
\usepackage{titletoc}
\usepackage{titlesec}
\usepackage[utf8]{inputenc}
\usepackage{enumerate}
\usepackage{hhline}

\makeatletter
\renewcommand{\Function}[2]{%
	\csname ALG@cmd@\ALG@L @Function\endcsname{#1}{#2}%
	\def\jayden@currentfunction{#1}%
}
\newcommand{\funclabel}[1]{%
	\@bsphack
	\protected@write\@auxout{}{%
		\string\newlabel{#1}{{\jayden@currentfunction}{\thepage}}%
	}%
	\@esphack
}
\makeatother

\usepackage[english]{babel}

\makeatletter
\def\BState{\State\hskip-\ALG@thistlm}
\makeatother

\title{Fast Stochastic Peer Selection in Proof-of-Stake Protocols}

\author{Quan Nguyen, Andre Cronje, Michael Kong}
\affil{FANTOM}

\begin{document}
\maketitle

\begin{abstract}
The problem of peer selection, which randomly selects a peer from a set, is commonplace in Proof-of-Stake (PoS) protocols. In PoS, peers are chosen randomly with probability proportional to the amount of stake that they possess.

This paper presents an approach that relates PoS peer selection to Roulette-wheel selection, which is frequently used in genetic and evolutionary algorithms or complex network modelling. 
In particular, we introduce the use of stochastic acceptance algorithm~\cite{lipowski2012roulette}
for fast peer selection. The roulette-wheel selection algorithm~\cite{lipowski2012roulette} achieves $O(1)$ complexity based on stochastic acceptance, whereas searching based algorithms may take $O(N)$ or $O(logN)$ complexity in a network of $N$ peers.

\end{abstract}



\section{Introduction}\label{ch:intro}

There have been an upsurge of interests in cryptocurrencies and distributed ledger technologies since Bitcoin's birth. The underlying blockchain technologies (BCT) have enabled numerous opportunities for business and innovation across different domains from financial, healthcare to logistics.

\emph{Proof Of Stake} (PoS): is a promising approach to overcome the limitations of PoW, such as electricity consumption and low confirmation rate. PoS~\cite{ppcoin12, dpos14, dagcoin15, algorand17, sompolinsky2016spectre, PHANTOM08} leverages participants' stakes for selecting the creator of the next block~\cite{ppcoin12,dpos14}. Validators have voting power proportional to their stake. PoS is considered more secure than PoW, as each participant is incentivised to maintain the sustainability of the network.

\emph{PoS+DAG} (PoS-DAG): StakeDag protocol~\cite{stakedag} and StairDag~\cite{stairdag} present a general model that integrates PoS model into DAG-based consensus protocols to aim for more safety and liveness of the network. Like its predecessors~\cite{fantom18}, both consensus protocols generate each block asynchronously to build the DAGs. Consensus on a block is computed from the validating power of Users and Validators of parent blocks. In both protocols~\cite{stakedag,stairdag}, a node randomly selects the next peer(s) to gossip with and to generate new event blocks, based on participants' stakes. 

\subsection{Peer Selection}\label{sec:peerselection}
Let us consider a network of $N$ peers with positive weights $w_i$'s. The random peer selection in PoS and PoS-DAG is the problem of selecting a peer randomly from the $N$ peers. The probability of selection of a peer $i$ is proportional to the weight $w_i$ of $i$.
In StakeDag and StairDag, our random peer selection aims not only to select highly trusted peers more frequently than others,
but also to prevent any single peer from dominating the network.

\section{Peer Selection as Roulette Wheel Selection}

\subsection{Roulette Wheel Selection}
Selection is a crucial part of genetic algorithms (GAs) as good selection will drive fitter solutions in next generation and will have a significant impact on their convergence. Yet it is also important to prevent one extremely fit solution from dominating the entire population as this leads to a loss of diversity.
As a simple rule, the fitter an individual, the higher chance of its survival and propagation of its features to next generation. 

A straight-forward implemention of the rule is Roulette-wheel selection~\cite{goldberg89}. The method assumes that every individual can be selected with probability proportional to its fitness. Let us consider $N$ individuals, each characterized by its fitness $w_i>0$ ($i$= 1,2, $\dots$ , $N$). 
The selection probability of the $i$-th individual is thus given as:
$
p_i = \frac{w_i} {\Sigma_{i=1}^{N} w_i}
$.

This is similar to a Roulette wheel in casino. A circular wheel consists of $N$ pies. Each piece gets a portion of the circle, which is proportional to its fitness value $w_i$. Then a random selection is made similar to how the roulette wheel is rotated.
Selection of an individual pie is then equivalent to choosing randomly a point on the wheel and determining the corresponding pie. For simple search algorithms, such a location requires $O(N)$ operations while the binary search needs $O(logN)$.

{\bf Stochastic Acceptance algorithm}: A fast algorithm for Roulette wheel seelction is introduced based on Stochastic Acceptance~\cite{lipowski2012roulette}. The algorithm has $O(1)$ complexity.

Let $w_{max}$= max $\{w_i\}^N_{i=1}$ is the maximal fitness in the population. The algorithm consists of the following steps:
\begin{enumerate}
\item Select randomly an individual (say, $i$) with uniform probability (1/$N$).
\item The selection is accepted with probability $\frac{w_i}{w_{max}}$ . If not accepted, the procedure is repeated from step 1.
\end{enumerate} 
Interested readers may see more details in the original paper~\cite{lipowski2012roulette}.

\subsection{An Application to Peer Selection}
The random peer selection problem, which is described in Section~\ref{sec:peerselection}, is equivalent to the Roulette wheel selection problem. The weight $w_i$ is the participants' stakes in PoS peer selection, and is equivalent to the fitness value in Roulette wheel selection.
The probability of selection of a peer in PoS is proportional to stake value $w_i$, whereas that probility of selection of an individual in GA is proportional to its fitness value. 
Like roulette selection, peer selection also aims to prevent loss of diversity and to avoid any single peer from dominating the population.

We have implemented the Stochastic Acceptance algorithm~\cite{lipowski2012roulette} in Go. Our initial experiments with the implementation have shown that the algorithm runs quite fast. The benchmarking result, depicted in Figure 3 of their paper~\cite{lipowski2012roulette}, shows that their algorithm is significantly faster than the search-based alternatives. The algorithm is promising to speed up peer selection in PoS systems.

\section{Conclusion}\label{se:con}
We have shown that the peer selection problem in Proof of Stake consensus protocols is equivalent to Roulette wheel selection problem. Further, we have proposed to use a fast Roulette wheel selection algorithm~\cite{lipowski2012roulette} to achieve $O(1)$ complexity, for peer selection in PoS systems.
The established connection will enable several potential directions for future work in blockchain technologies.

\section{Reference}\label{se:ref}

\renewcommand\refname{\vskip -1cm}
\bibliographystyle{abbrv}
\bibliography{LCA}

\end{document}